\documentstyle[galley,epsfig]{mn2e}
\title{Abundance analysis of  s-process enhanced   
barium stars}
\title[Abundance  analysis of  s-process enhanced  barium stars]
{Abundance  analysis of  s-process enhanced  barium stars  }

\author[Upakul Mahanta et al.]{Upakul Mahanta$^{1,2}$, Drisya Karinkuzhi$^{3}$, Aruna Goswami$^{3}$, Kalpana Duorah$^{2}$  \\
    $^{1}$ Department of Physics, Bajali College, Pathsala, Assam,
781325, India; upakulmahanta@gmail.com  \\
    $^{2}$ Department of Physics, Gauhati University, Guwahati
781014, Assam, India; khowang56@yahoo.co.in\\
    $^{3}$Indian Institute of Astrophysics, Koramangala, Bangalore
560034, India; \\
drisya@iiap.res.in, aruna@iiap.res.in\\
}

\begin{document}

\date{ Accepted ---;  Received ---;  in original form : }

\pagerange{\pageref{firstpage}--\pageref{lastpage}} \pubyear{2016}

\maketitle

\begin{abstract}
Detailed chemical composition studies of stars with  enhanced 
abundances of neutron-capture elements  can provide observational 
constraints for neutron-capture nucleosynthesis studies and clues 
for  understanding their contribution to the Galactic chemical 
enrichment. We present   abundance results from high-resolution 
spectral analyses of  a  sample of four chemically peculiar stars 
characterized by  s-process enhancement. High-Resolution spectra 
(R $\sim {42000}$) of these objects  spanning a wavelength range  
from $4000$ to $6800$ \r{A}, are taken from the ELODIE archive. We 
have estimated the stellar atmospheric parameters, the effective 
temperature T$_{eff}$, the surface gravity log g, and metallicity 
[Fe/H] from local thermodynamic equilibrium analysis using model 
atmospheres. We report  estimates of elemental abundances for several 
neutron-capture elements, Sr, Y, Zr, Ba, La, Ce, Pr, Nd, Sm, Eu and Dy.     
While HD~49641 and HD~58368 show [Ba/Fe] ${\ge}$ 1.16 the other two 
objects HD~119650 and HD~191010  are found to be mild barium stars 
with [Ba/Fe] ${\sim}$ 0.4. The derived abundances of the  elements are 
interpreted on the basis of existing theories for understanding their 
origin and evolution.

\end{abstract}


\section{Introduction}
 
The origin and evolution of neutron-capture elements in our Galaxy 
 still remain  poorly  understood. Detailed chemical composition of stars
with  atmospheres enriched with heavy elements can provide clues for 
better understanding.  Barium stars along with CH stars
and the more metal-poor analogs
of CH stars, the  Carbon-Enhanced Metal-Poor (CEMP)-s stars provide
important targets.
Barium stars, first identified as a distinct group of peculiar objects 
by Bidelman \& Keenan (1951) are Pop I red gaints with spectral class G to K. 
Their spectra show strong presence of singly ionized Ba II 
lines, indicating an overabundance of s-process elements. 
Variations in radial-velocity  in barium stars   was revealed by 
McClure et al. (1980) and  further confirmed by McClure (1983, 1984) 
and Udry et al. (1998a,b).  The binary nature of barium stars
was established from the observed radial velocity variations 
that suggested   presence of companions. The most accepted  
interpretation for the  observed  abundance patterns in barium stars is that, 
they acquired the enrichment of s-process nucleosynthesis  from an 
AGB companion that later evolved to a white dwarf. Detailed 
abundance studies of barium  stars are thus important  for a better 
understanding of the s-process mechanisms in AGB stars, as well as 
  the mass transfer mechanisms in binary systems. 

Present theories include two main 
types of mass transfer mechanisms for barium  stars in binary systems, the 
Roche-lobe overflow (RLOF) and wind-accretion scenarios. 
 In the  RLOF scenario, for sufficiently close binaries
containing a  thermally pulsing AGB (TP-AGB) star, as the star 
expands past the Roche lobe the material which lies outside the 
lobe can fall off into the secondary object.  In the wind accretion 
model of  Boffin \& Jorrisen (1988), the authors have shown that 
for large orbital separation the contamination of Ba stars 
depends only upon the present value of orbital separation and 
mass of the white dwarf companion. Here part of matter are ejected 
through a wind by an AGB star and the system remain always detached. 
This is the more favoured channel for barium star formation.   
 An intermediate regime in which some wind mass-transfer occurs 
before RLOF begins  has also been suggested (Han et al. $1995$). 
However,  none of these models can explain all the physical and 
orbital properties of Ba stars. While RLOF scenario is not 
favourable as this terminates the TP-AGB and lessens the production 
of barium, preventing barium-star formation,  the wind-accretion 
scenario does not explain the distribution of eccentricities and 
periods of barium stars (Izzard et al. $2010$). 
Further studies are necessary to have clearer pictures on these issues.
As much insight about the origin and evolution of  heavy elements 
and the  evolved AGB companion can be obtained 
from the chemical composition studies of these objects, 
and, given the current   number of abundance analyses  available,
(although   gradually increasing  but  for limited
number of heavy elements), detailed abundance studies of even one 
or two more objects represent a significant contribution.

In the present  work we have undertaken to  conduct a detailed chemical 
composition study for  four peculiar stars listed in the barium 
stars catalogue of L\"u (1991), based on high resolution 
spectra (R $\sim$ 42000) and high S/N with $>$ 100 at about 5500 \AA\,.  
 Among these four stars two
objects HD~49641 and HD~58368 are included in the list of  `Certain 
Ba II stars' and one object HD~119650 is included in the list of
`Marginal Ba II stars' of MacConnell et al. (1972). No detailed studies 
were available in literature for these objects until recently when
these  two  objects, HD~49641 and HD~58368 are  found to be   included 
in  the chemical composition studies of barium stars by Yang et al. (2016) 
and de Castro et al. (2016). 
As far as the heavy element abundances are concerned,  Yang et al. (2016) 
have reported abundances for  five heavy elements
(Y, Zr, Ba, La and Eu); likewise de Castro et al. (2016) have also
reported on abundances of five heavy elements (Y, Zr, La, Ce, Nd). 
We have presented abundances for eleven neutron-capture elements and 
compared and contrast  our results for these two objects with the 
results of the above two studies whenever possible.
The other two objects, hitherto not studied, are found to be
mild barium stars  with [Ba/Fe] ${\sim}$ 0.4.
 
In section $2$ we have presented a brief summary on the chemical composition
 studies on HD~49641 and HD~58368 by Yang et al. (2016) and 
de Castro et al. (2016). 
In section $3$ we have presented  the details of  the high resolution 
spectra used in this study.  Photometric temperatures of the stars  
and the radial velocity estimates are also presented  in this section.
  Section $4$ describes
briefly the methodology used  for deriving the stellar atmospheric 
parameters.  Error analysis is presented in section $5$. Estimation of 
chemical abundances and the analysis of the 
abundance results are discussed  in Section $6$. 
 Conclusions are presented  in Section $7$.

\section{Previous studies on HD~49641 and HD~58368: A summary}
Some aspects of these two objects were addressed by Yang et al. (2016) and
de Castro et al. (2016); here we summarize their main results.

Yang et al. (2016): These two objects were included in their 
chemical abundance studies of 19 barium stars. They have determined
the stellar atmospheric parameters using red spectra in the wavelength
range 5500 - 9000 \AA\, with a resolution R ${\sim}$ 30,000 and S/N ${\ge}$ 60.
The derived T$_{eff}$ obtained using color indices (B $-$ V) and the
empirical calibration of Alonso et al. (1999, 2001) for giant stars are
lower than our estimates  by 350 K in case of HD~49641 and by 318 K
in case of HD~58368. Their lower values are likely  due to the fact that
the color (B $-$ V) is affected by CN and C$_{2}$, making the stars redder
and as a result the temperature derived from this color may be lower
(Allen And Barbuy 2006). Estimated log g obtained by them  using Hipparcos
parallexes  also differ significantly from our estimates. 
The chemical abundance estimates  of 
 Na, Al, ${\alpha}$- and iron-peak elements (O, Na, Mg, Al, Si, Ca,
Sc, Ti, V, Cr, Mn, Ni) of Yang et al.  are  found to be  similar to 
the solar abundances.  Overabundances
of neutron-capture  process elements relative to the Sun are  
noticed in both the stars. 
The Y I and Zr I abundances are lower
than Ba, La and Eu, but higher than the ${\alpha}$- and iron-peak elements.
 The authors noted   a positive correlation 
between Ba intensity and [Ba/Fe] in their sample of stars.
For the n-capture elements (Y, Zr, Ba, La), there is an anti-correlation 
between their [X/Fe] and [Fe/H]. With [Ba/Fe] = 1.13 and 0.98  respectively  
for HD~49641 and HD~58368, they have confirmed these  objects to be
strongly enhanced barium stars.

de Castro et al. (2016): In this work the authors have presented an 
homogeneous analysis of photospheric abundances based on
high-resolution spectroscopy of a sample of 182 barium stars and candidates.
HD~49641 and HD~58368 are two members of this extended sample. 
The methodology employed by these authors to derive the stellar parameters
are similar to ours. However, 
there lies  significant differences between ours and their estimates
of stellar atmospheric parameters, T$_{eff}$ and surface gravity log g.
While de Castro et al's estimates for temperature and log g for HD~49641 are
 4400 K and 1.5 respectively our estimates are significanltly 
higher with values  4700 K and 3.4. 
Their estimates of  the  abundances of ${\alpha}$-elements and iron peak 
elements  are  similar to those of field giants with the same metallicity. 
The observed enrichment in sodium 
is  attributed to the operation of Ne-Na cycle. These two
stars are also found to exhibit  overabundance of the elements created by 
the s-process.
From the  measurements of the mean heavy-element abundance pattern as 
given by the  ratio [s/Fe], it was found that the barium
stars present several degrees of enrichment. 
From the measured 
photospheric abundances of the Ba-peak and the Zr-peak elements, 
 [s/Fe] and the [hs/ls] ratios (where hs refers to 
heavy s-process elements and ls refers to light s-process elements) 
are found to be strongly 
anticorrelated with the metallicity, and that, the 
barium stars  follow an age–metallicity relation. 

A  comparison   of our estimated  elemental abundance  ratios with their 
counterparts in Yang et al. (2016) and de Castro et al. (2016) is 
presented  in later sections. 

\section{High Resolution Spectra of the sample stars }
High resolution spectra (R$\sim{42000}$) of the four objects 
selected from the Barium Star Catalogue (L\"u $1991$) are taken 
from the ELODIE archive (Moultaka et al. (2004)). The  echelle 
spectrograph ELODIE (Baranne et al. 1996) is attached to the  
$1.93$-m telescope of the Observatoire de Haute Provence (OHP). 
The spectra  cover a  wavelength range from $4000$ to $6800$  \r{A}.  
The S/N ratios of these spectra are in the range 109 to 149 at 5500 \AA\,. 
An online reduction software {\footnotesize{TACOS}} 
automatically performs an optimal  extraction and wavelength 
calibration of the data. A spectrum is  recorded in a single exposure 
on a $1$K CCD that has  $67$ orders. 
The basic data for these objects are listed in Table 1.  
A few sample spectra are shown  in figure 1.

\begin{figure}
\centering
\includegraphics[height=8cm,width=8cm]{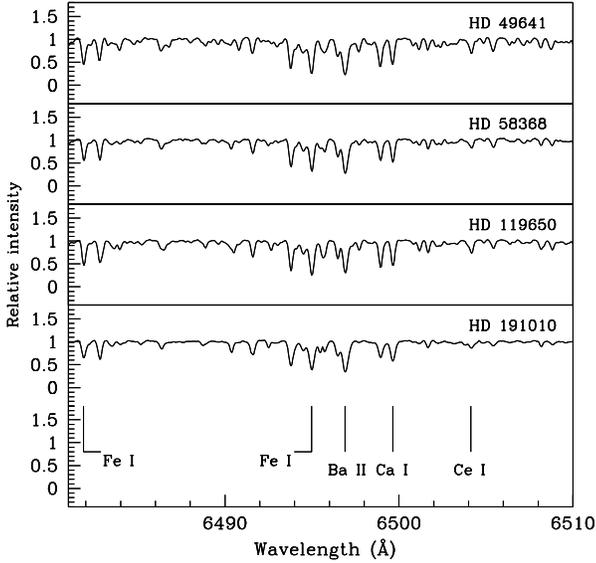}
\caption{ Sample spectra of a few programme stars in the  wavelength 
region 6480 to 6510 {\bf  {\rm \AA}}.}
\end{figure}

{\footnotesize
\begin{table*}
{\bf Table 1: Basic data for the programme stars}\\ 
\begin{tabular}{lccccccc}
\hline
Star ID     &RA$(2000)$       &Dec.$(2000)$    &B       &V       &J        &H        &K  \\
\hline
HD 49641    &06 49 29.4495    &+03 41 30.189   &8.47  &7.12  &5.078  &4.506  &4.253\\
HD 58368    &07 25 38.9681    &+07 33 39.046   &9.01  &7.99  &6.334  &5.919  &5.773\\
HD 119650   &13 44 27.1157    &+00 42 07.020   &8.78 &7.59  &5.668  &5.164  &4.908\\
HD 191010   &20 06 45.8410    &+25 41 06.790   &9.13  &8.17  &6.488  &6.092  &5.970\\

\hline

\end{tabular}
\end{table*}
}

\subsection{Photometric temperatures}
Photometric temperatures for  the objects  presented in Table 2 are
estimated using the colour temeprature  calibrations  from Alonso 
et al. (1999) (i.e.,  equations 4, 9, 10, 11 in Table 2) with its 
erratum in  Alonso et al. (2001). 
Necessary transformations between the different  photometric systems, i.e., 
from  Ramirez and Melendez (2004) and Alonso et al. (1996, 1999) are used 
to account for the difference between 2MASS infrared photometric     
system and photometry data measured on the TCS system  used in 
Alonso et al. calibrations.

{\footnotesize
\begin{table*}
\begin{tiny}
{\bf Table 2: Temperatures from  photometry }\\
\begin{tabular}{llllllllllll}
\hline
Star Name  &  $T_{eff}$  &  $T_{eff}$  &  $T_{eff}$  & $T_{eff}$  & $T_{eff}$  & T$_{eff}$  &T$_{eff}$  & $T_{eff}$  &  $T_{eff} $  &  $T_{eff}$    & Spectroscopic   \\
          &             &  $(-0.5)$    &   $(-0.5)$  &  $(-1.0)$  & $(-1.0)$   & $(-1.5)$ &
$(-1.5)$  & $(-0.5)$    &  $(-1.0)$    & $(-1.5)$    & estimates  \\
           &  (J-K)      &   (J-H)          &   (V-K)          &  (J-H)          &    (V-K) &  (J-H) & (V-K) &(B-V) & (B-V) & (B-V) &  \\
\hline

HD49641   &  4199  &   4526 &   4312 &   4544 &   4298  &   4534 &   4291&3919 &  3686&  3497& 4700\\
HD58368   &  4918  &   5135 &   4863 &   5163 &   4851 &    5154 &   4846&4433 &  4167 &  3953 &5095 \\
HD119650  &  4350 &   4779 &   4445 &   4802 &   4432  &   4792 &   4425&4152 &  3904 &  3703& 4825  \\
HD191010  &  5068  &   5219 &   4881 &   5248 &   4869  &   5240 &   4864&4541  & 4268 &  4049&5325 \\
\hline
\end{tabular}
\end{tiny}

The numbers in the parenthesis below $T_{eff}$ indicate the metallicity values at which 
the temperatures are calculated. Temperatures are given in Kelvin\\
\end{table*}
}

\subsection{Radial velocity}
A  set of clean unblended lines from each of the object spectrum was
selected    
 to estimate the radial velocities. The estimated mean radial 
velocities are presented in Table 3 along with the literature values 
for a comparison. 
The objects show significant variations in radial velocities.

{\footnotesize
\begin{table*}
{\bf Table 3. Radial Velocities of programme stars}\\
\centering
\begin{tabular}{|c|c|c|c|c|c|}
\hline
Object    &V$_r$ km s$^{-1}$&S/N  of our& Date of  &V$_r$ km s$^{-1}$ & References  \\
          &(our estimates) &  spectra at 5500 \AA\,  & Observation&     (literature values)&    \\
\hline
HD49641   &   7.20 &  121  &17.12.2000 & 4.45  &  1     \\
HD58368   &  41.67 &  109  &26.11.2001 & 37.81  & 1     \\
HD119650  &  -5.10 &  143  &23.5.2003  &-5.70  & 2     \\
HD191010  &  20.60 &  149  & 29.7.2003 & 18.10  & 2     \\ 
\hline
\end{tabular}

1. Pourbaix et al. (2004); 2. Gontcharov (2006)
\end{table*}
}

\section{STELLAR ATMOSPHERIC PARAMETERS}
 A selected  set of unblended Fe I and Fe II lines  (Table 4) with excitation 
potential in the range 0.0 - 5.0 eV and 
equivalent width of 20 to 190 m\r{A} is considered for the determination of
the stellar parameters. 
We have assumed local thermodynamic equilibrium (LTE) for the analysis. 
An updated 2014 version of  MOOG (Sneden 1973)  was used for 
the calculations. The model atmospheres were taken 
from the Kurucz grid of model atmospheres with no convective overshooting.
Solar abundances of the elements are adopted from  Asplund, Grevesse \& 
Sauval (2009). 

The microturbulent velocity was determined  at a given effective temperature 
by forcing  that there be no dependence of the abundances determined using 
 Fe I lines  on the equivalent width of the corresponding lines.
The effective temperature was determined by making the slope of the 
abundance versus the excitation potential of Fe I lines to be 
nearly zero. The initial value of the temperature was taken 
from the photometric estimates, and a final value arrived at by an 
iterative method.   A few examples for   
 the  observed abundances  of Fe I and Fe II as a function of excitation 
potentials  and equivalent widths are shown in figure 2.
The surface gravity was fixed at a value that gives the same 
abundances for Fe I and Fe II lines. 
Clean Fe II lines are more difficult to find than Fe I lines. In most cases,
 we were limited to 4 - 10 Fe II lines for our  analyses.
   Derived atmospheric parameters are listed in Table 5.

\begin{figure}
\centering
\includegraphics[height=10cm,width=9cm]{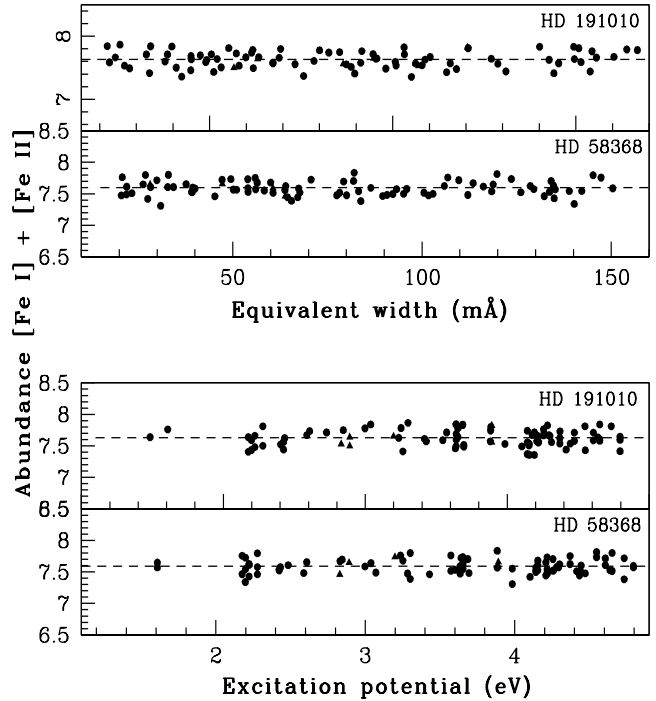}
\caption{The iron abundances of stars  are shown for
individual Fe I and Fe II lines as a function of excitation potential 
in the lower panel. Similarly 
the iron abundances of stars  are shown for individual Fe I 
and Fe II lines as a function of equivalent width in the upper panel.The
abundances correspond to the adopted atmospheric parameters of the stars. 
The solid circles indicate Fe I lines and solid triangles  indicate 
Fe II lines.}
\end{figure}

\begin{figure}
\centering
\includegraphics[height=10cm,width=9cm]{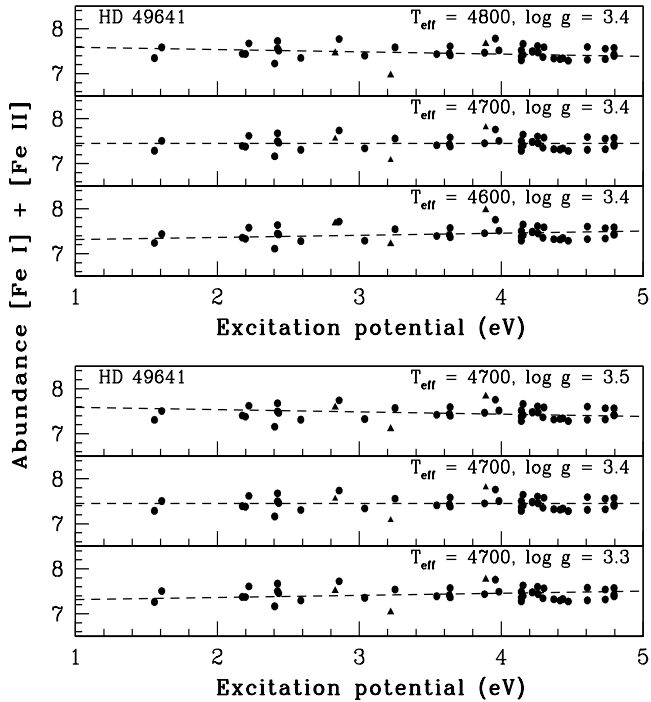}
\caption{The iron abundances of stars are shown for
individual Fe I and Fe II lines as a function of excitation potential.
The solid circles indicate Fe I lines and solid triangles  indicate 
Fe II lines. In the upper panel the middle plot  shows the abundances 
corresponding to the 
adopted T$_{eff}$. In the same panel  the upper and the lower plots 
show the variation in the
abundance trend due to the variation of T$_{eff}$ by 
$+$ 100 K and $-$ 100 K respectively. In the lower panel the middle plot
 shows the abundances corresponding to the
adopted log g. In the same panel  the upper and the lower plots show 
the variation in the
abundance trend due to the variation of log g by 
$+$ 0.1 and $-$0.1 respectively. }
\end{figure}

\section{ERROR ANALYSIS}
 
The errors in the elemental abundances are mainly due to the 
errors in deriving  the atmospheric parameters which have contributions 
from the measurement of the equivalent widths of Fe lines. 
The minimum error 
in the elemental abundances  are taken as the standard deviation of 
the Fe abundance derived for each objects. These are   listed in 
Tables 6 and 7. 
The minimum error in temperature is taken as ${\pm}$ 100 K and for log g as 
${\pm}$  0.1 dex. 
We have calculated the errors in the atmospheric parameters as described 
by Ryan et al. (1996). 

In order to show the effect of temperature variations  in the abundance 
pattern we have plotted the 
Fe abundances with respect to the  excitation potential values at three 
different temperature values, the adopted temperature (T$_{eff}$)  and  
T$_{eff}$ $\pm$ 100 K (figure 3, upper panel). Similarly, the effect on 
abundance trends
due to variation in the adopted log g values by $\pm$ 0.1 dex is also shown   
in figure  3 (lower panel). The errors presented in tables 6 and 7 
correspond to the 
standard errors when abundances are measured using more than one line.
For abundances that are derived using a single line or by using spectrum
synthesis calculations an error of $\pm$ 0.2 dex is considered. A
change by an amount $\pm$ 0.2 dex from the adopted abundance is 
found to be  necessary to notice a  visual separation between the  
different synthesized plots. 

{\footnotesize
\begin{table*}
{\bf Table 4: Equivalent widths (in m\r{A}) of Fe lines used for deriving 
atmospheric parameters.}\\ 
\begin{tabular}{ccccccccc}
\hline
Wavelength  &id         &E$_{low}$ & log gf   &  HD 49641      	& HD 58368 	& HD 119650	 &HD 191010 &References    \\
(\r{A})     &           & (eV)     &          &           	&          	&          	 &       	&\\\hline 
 4348.937   &    Fe I	&   2.990  &  -2.13  &      --   	&    --    	&   103.9(7.50) &     --    &1	\\
 4551.649   &    	&   3.943  &  -2.06  &      --   	&    --   	&    47.8(7.30) &     --   &1	\\
 4566.514   &    	&   3.301  &  -2.25  &      --   	&  65.5(7.39)  &     --   	 &     --   &1	\\
 4587.716   &    	&   3.984  &  -2.15  &      --         &  30.9(7.31)  &     --   	 &     --  &1	\\
 4602.000   &    	&   1.608  &  -3.15  &       --  	&  118.8(7.65)  &  129.2(7.38)  &     --  &1	\\
 4625.044   &   	&   3.241  &  -1.34  &       --  	&    --		&  --            &     --     &1    \\
 4741.529   &   	&   2.831  &  -2.00  &       --  	&   113.5(7.67) &  122.8(7.48)  &     --  &1	\\
 4745.800   &   	&   3.654  &  -0.79  &       --  	&   134.7(7.63) &  --            &     --  &1	\\
 4787.827   &   	&   2.998  &  -2.77  &       --  	&    67.4(7.59)&     --  	 &   80.0(7.78)   &2\\ 
 4788.751   &  		&   3.236  &  -1.81  &       --        &   106.8(7.76) &  --            &  108.8(7.63)   &1 \\
\hline
\end{tabular}

The numbers in the paranthesis in columns 5-8 give the derived abundances from the respective line.\\. 
References: 1. F\"uhr et al. (1988) 2. Kurucz (1988)\\
{\bf Note:} This table is available in its entirety in online only. A portion is shown here for guidance regarding its form and content.\\
\end{table*}
}

{\footnotesize
\begin{table*}
{\bf Table 5: Derived atmospheric parameters for the programme stars. }\\ 
\begin{tabular}{cccccc}
\hline
Star ID   & T$_{eff}$  & log g   &$\zeta$          & [Fe I/H]  &[Fe II/H]\\
          &    (K)     &         & (km s$^{-1}$)    &           &          \\
\hline
HD 49641  &  4700      &  3.40   & 1.77            &   -0.05    & -0.02    \\
HD 58368  &  5095      & 3.45    & 1.37            &   0.09    & 0.12    \\
HD 119650 &  4825      & 2.85    & 1.62            &   0.04    & 0.14 \\
HD 191010 &  5325      & 2.38    & 1.92            &   0.13    & 0.12   \\
\hline
\end{tabular}
\end{table*}
}

\section{Abundance analysis}
Elemental abundances are  calculated  from  the measured  equivalent
 widths of lines due to neutral and ionized elements using the 2014 
version of MOOG (Sneden 1973) and  the adopted model atmospheres. 
From a close comparison of the spectra of the programme stars with that
of the spectrum of Arcturus a line list of all the elements  was generated. 
However, only the  lines that were used for abundance analysis are included 
in the line list; others being  blended with contributions from other 
species could not be used for abundance analysis.
The primary source of  the adopted  log gf 
values  for the atomic lines due to  these elements   is Kurucz 
atomic line database (Kurucz 1995a, b);  other sources from literature,
 such as,   Karinkuzhi \& Goswami (2014, 2015), Goswami \& Aoki (2010),  
Aoki et al. (2005, 2007), Goswami et al. (2006, 2016), 
Jonsell et al. (2006),  Sneden et al. (1996), Luck and Bond (1991) 
 were also consulted. 
For the elements Sc, V, Mn, Ba, La and Eu, spectrum synthesis
calculation considering hyperfine structure  is used to find the abundances. 
The line list for each region  synthesised is taken from Kurucz 
atomic line list (http://www.cfa.harvard.edu/amdata/ampdata/kurucz23/\\
sekur.html).  For a  few La lines the log gf values 
are taken from Lawler et al. (2001). 

We had also  estimated the
Fe abundances considering  our line list (Table 4) and using stellar 
atmospheric models corresponding to  the stellar parameters of 
 Yang et al. (2016)
 for the two objects HD~49641 and HD~58368.  In either case we 
did not arrive at the desired zero slope for `abundance vs excitation potential'
plot  and `abundance vs equivalent width' plot. 
We have  therefore  adopted the stellar parameters
estimated by us in determining   the abundances of the elements
presented in Tables 6 and 7. 
 Derived abundance values along with the abundance ratios with respect 
to iron are listed in Tables 6 and 7. 
Lines used for the abundance calculation of these elements are listed in 
Table 8. 

To examine the accuracy of our estimates  we have also measured on 
solar spectra  the equivalent widths for a large number of lines 
(these lines are also detected in the spectra of our programme
stars) and measured the solar abundances using the log gf values
that we have adopted. The solar atmospheric parameters used are
microturbulent velocity 1.25 km s$^{-1}$,  T$_{eff}$ = 5835 K and 
 log g = 4.55 cm s$^{-2}$.  When comparing with Asplund et al. (2009), the 
 derived abundances are found to  match closely within a
range of 0.08 $-$ 0.1 dex.
A comparison
of our estimated  abundance ratios for elements Na to Zn with respect to Fe, 
with their  counterparts in normal giants
 and barium stars obtained from literature are shown in figure 4. 
In figure 5,  we have illustrated
a few examples of spectrum synthesis calculations for Y, Ba and La.
In Table 9, we have presented [ls/Fe], [hs/Fe] and [hs/ls] values,
where ls represents the light s-process elements Sr, Y and Zr and hs
represents the heavy s-process elements Ba, La, Ce, Nd and Sm. A comparison 
of our estimated  abundance ratios for neutron-capture elements 
with their counterparts in  normal giants and 
  barium stars obtained from literature are shown in figure 6.

\begin{figure}
\centering
\includegraphics[height=8cm,width=8cm]{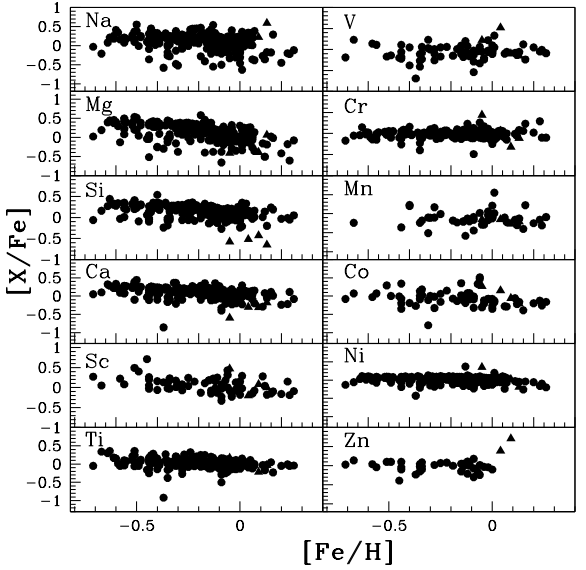}
\includegraphics[height=8cm,width=8cm]{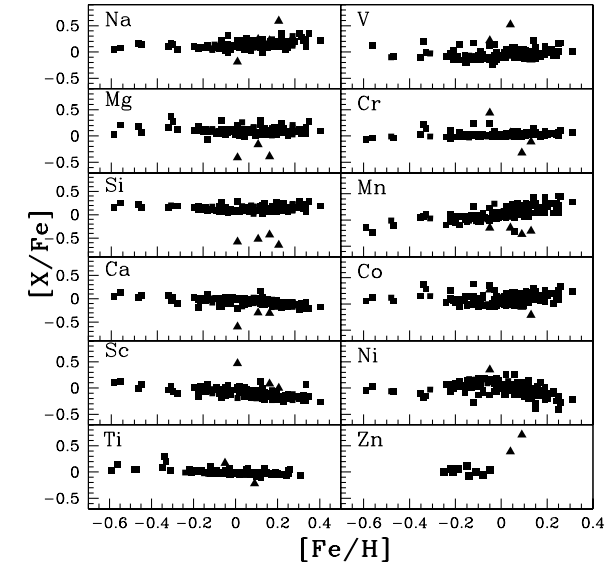}
\caption{Abundance ratios of light elements observed in the programme
stars with respect to [Fe/H]. In the upper  panel our program stars (solid 
triangles) are compared with the abundance ratios observed in  Ba stars
(solid circles) from literature (i,e., Allen \& Barbuy (2006),
 Liang et al. (2003), Smiljanic et al. (2007), Zacs (1994), de Castro 
et al. 2016)).  In the lower panel our program stars 
 (solid triangles) are compared with the abundance ratios observed
in giants (solid squares) from literature (i.e., Luck and Heiter (2007), 
Mishenina et al. (2006)). 
}
\end{figure}

\begin{figure}
\centering 
\includegraphics[height=8cm,width=8cm]{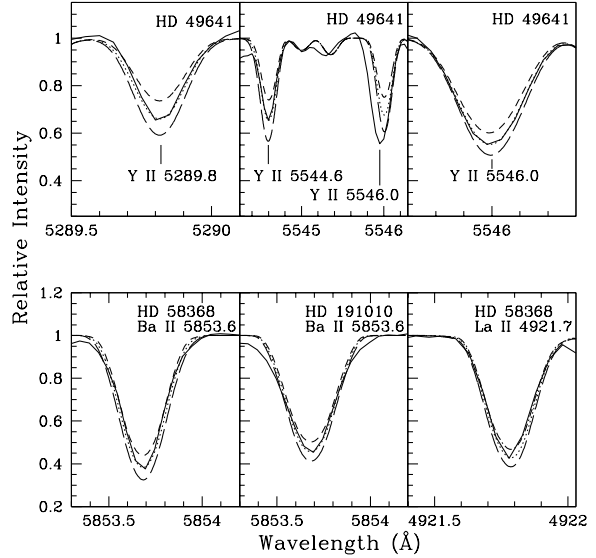}
\caption{Spectral-synthesis fits of Ba II line at 5853.6 {\rm \AA} and 
La II line  at 4921.7 {\rm \AA} are shown for two objects HD 58368 and 
HD 191010 in the lower panel. 
In the upper panel, spectral-synthesis fits of Y II line at 5289.815 {\rm \AA},
5544.615 {\rm \AA}, and 5546.009 {\rm \AA}, are shown for HD~49641.   
The dotted lines indicate the synthesized spectra and the solid
lines indicate the observed line profiles. Two alternative synthetic spectra
at the  (adopted value + 0.3, long-dashed line) and 
 (adopted value $-$ 0.3, short-dashed line) are shown to demonstrate 
the sensitivity of the line strength to the abundances.  we could get good fits
using the adopted value of the corresponding elements (Tables 6 and 7).}
\end{figure}
\subsection{Na and Al}

We have derived the Na abundance for all the programme stars using the 
lines at 5682.65 {\rm \AA} and 5688.22 {\rm \AA} whenever available.  
 The object HD~191010 shows a marginally higher abundance
ratio with [Na/Fe] = 0.59 than those observed in normal giants
(figure 4, upper panel). Many barium stars
studied by de Castro et al. (2016) also show sodium abundance 
ratios [Na/Fe] in the range 0.3 to 0.6. 
Sodium  does not show any trend with metallicities in field giants. Two of our 
objects HD~58368 and HD~119650 show sodium abundances similar to many
field giants with [Na/Fe] values  0.22,  and 0.24 respectively.  
The object HD~49641 however shows a mild underabundance with [Na/Fe] = $-$0.19
which is somewhat similar to that obtained for the object  HD~203137 
([Na/Fe] = $-$0.14) as  reported by Yang et al. (2016).
 
We could not estimate  Al abundance as no clear Al lines
could be detected due to line blending.

\subsection{ Mg, Si, Ca, Sc, Ti, V}
${\alpha}$-elements in field giants are known to show a slight increase 
with decreasing metallicity.
 Three stars in our sample (tables 6 and 7)  exhibit abundance ratios of Mg, 
and Ca lower than that generally seen in normal giants. Our estimated values 
are  however well within the range normally  seen in barium stars. 
This 
is not surprising as some barium stars are known to exhibit underabundance
(Yang et al.(2016) estimated a ratio of [Mg/Fe] ${\sim}$  $-$0.28 for 
the two barium  stars HD~31308 and
HD~224276).  While HD~191010 is found to show a near 
solar value our estimates for the other three objects are $-$0.41, $-$0.39 and 
$-$0.16 respectively for HD~49641, HD~58368 and HD~119650. 
 Yang et al. (2016) and de Castro et al. (2016) estimated [Mg/Fe] = $-$0.13 and $-$0.03   respectively for  HD~58368. 
Abundance ratios of Ti with respect to  iron  are similar
to those generally observed in normal giants and barium stars. 

In all the four stars Si abundance is found to be lower than what is
normally observed in barium stars and normal giants. Further studies 
including theoretical aspects  would  be worthwhile to understand these
anomalies. 

While Sc abundance ratios in three stars  are similar to those observed 
in normal giants, HD~49641 shows a higher value with [Sc/Fe] = 0.47. 
Sc abundance ratios are well within the range seen in 
 barium stars. The abundance ratio of V in HD~119650 is marginally
higher  ([V/Fe] = 0.52) than those seen in normal giants. Although 
abundance of V is calculated
using  measured equivalent widths for many lines, this being an odd Z element
its abundance is also estimated from  spectrum synthesis calculation 
of V I line 
at 5727.028 ${\rm \AA}$ taking into account the hyperfine components
from Kurucz database. The  values obtained from spectrum synthesis 
calculations are listed in the abundance tables. The  
Sc abundance is estimated using the 
Sc II line at 6245.63 {\rm \AA} considering hyperfine structure from 
Prochaska  \& McWilliam (2000). 

\subsection{Cr, Mn, Co, Ni, Zn}
Abundance ratios of iron peak elements  exhibit  similar values as seen in 
barium stars.  However, a close agreement with the values of normal
giants are not observed in all the cases. 
 Cr  abundances derived using Cr I lines show near solar values for 
HD~119650 (+0.06) and HD~191010 ($-$0.11). Cr is overabundant in HD~49641 
with [Cr/Fe] =  0.44 and  underabundant in HD~58368 with [Cr/Fe] = $-$0.32.
 Cr abundances measured using Cr II lines 
whenever possible also give  similar results.

Mn abundance is obtained using spectrum synthesis calculation of 
6013.51 ${\rm \AA}$ line taking into account the  hyperfine structures  from 
Prochaska $\&$ McWilliam (2000).

We could estimate Zn abundance only in two stars, 
  HD~58368 and HD 119650 using a single Zn I line at 4722.15 \AA\,. 
This line returns a value [Zn/Fe] = 0.71 and 0.39 respectively which
are higher than those normally seen in normal giants and barium stars
(figure 4). 

\subsection{ Sr, Y, Zr}
We have estimated the abundance of Sr  using the  Sr I line at 
4607.327 {\rm \AA}.  Sr is overabundant in all the four stars 
 with  [Sr/Fe] values  0.97,  1.19, 1.01 and 1.36 for HD~49641, 
HD~58368, HD~119650 and HD~191010 respectively.  None of 
the Sr II lines detected in the spectra   are found 
suitable for abundance estimate.

Among the three light s-process elements  Yang et al. (2016) 
and de Castro et al. (2016) have reported abundances only for Y and Zr.
These values for HD~49641 and HD~58368 are respectively 
(Y: 0.35, 0.41; Zr: 0.41, 0.25; Yang et al.)  and 
(Y: 0.89, 0.85; Zr: 0.53, 0.60; de Castro et al.). 

We could measure the abundance of Y  in all the four stars. 
Y is found to be significantly overabundant in HD~49641. We have performed
spectrum synthesis calculation for three Y II lines  
 at 5289.815, 5544.615 and 5546.009 
for the object HD 49641 (figure 5). An Y abundance of 3.50 gives satisfactory
fits for the lines of Y II  at 5289.815 and  5544.615 \AA\, giving 
[Y/Fe] = 1.31.  In  Table 6, we have listed this value.
The Y II  line at 5546.009 \AA\, can be fitted with a slightly higher 
abundance  4.15 that gives [Y/Fe] = 1.96. However, this line is blended
with a line due to vanadium.
Estimated  [Y/Fe] values are  1.12, 0.68 and 0.11
in  HD 58368, HD~119650 and  HD~191010  respectively.

We could derive Zr abundance for three objects from Zr I lines.  
HD~49641 and HD~58368 both show an  overabundance with [Zr/Fe] =  0.73.   
 HD~119650 also show a mild 
enhancement with a value of  [Zr/Fe] = 0.42. Zr abundance could not be
estimated for HD~191010.

\subsection{ Ba, La, Ce, Pr, Nd, Sm, Eu, Dy}
Abundances ratios of these elements with respect to iron are compared with
their counterparts observed in normal giants and barium stars in figure 6.
We could estimate abundances of   these elements in all the  stars except 
for HD~49641 where abundances of Sm and Dy could not be estimated. While
de Castro et al (2016) had estimated abundances of only La, Ce and Nd,
for HD~ 49641 and HD~58368, Yang et al. (2016) estimated abundances of
only Ba, La, and  Eu for these two objects.   

As can be seen from Table 12, for a few elements our estimates  
are significantly
different from these two studies.  As expected,  the two stars with 
[Ba/Fe] $>$ 1.0,  show  abundances  for other heavy elements higher values 
than that generally noticed in
normal giants.  The object HD~119650
 with [Ba/Fe] $<$ 0.52 show [X/Fe] $>$ 1.0 for La, Sm and Dy and for the
rest of the  elements, [X/Fe] $<$ 1.0. In case of HD~191010  the heavy 
elements are found to  show [X/Fe] $<$ 0.52.  The  heavy elements 
in normal giants and barium stars  are known
to show large scatter with respect to metallicity.
 
As many lines due to Ce, Pr, Nd, and Sm  could be measured on our 
spectra, the standard abundance determination method using equivalent 
width measurements are used for  abundance estimates. 
Spectrum synthesis calculation is  performed  for Ba, La and Eu. 
The presented abundance for Ba, La and Eu in the Tables 6 and 7 are from 
spectrum synthesis calculations and  indicated in  Table 8. 

 Ba abundance is derived from  spectrum  synthesis  
calculation using Ba II line at 5853.668 ${\rm \AA}$ considering hyperfine 
components from McWilliam (1998). 
 La abundance for all the  stars are derived from 
spectrum  synthesis calculation of La II line at 4921.77 ${\rm \AA}$ 
considering  hyperfine components from Jonsell et al. (2006), 
and,   Eu abundance is determined  
from  spectrum synthesis calculation of Eu II line at 6645.130 ${\rm \AA}$ by 
considering  the hyperfine components from  Worley et al. (2013).

{\footnotesize
\begin{table*}
{\bf Table 6 : Elemental abundances in HD~49641 and HD~58368}\\
\begin{tabular}{|c|c|c|c|c|c|c|c|c|}
\hline
& \multicolumn{6}{c}{HD~49641} & \multicolumn{1}{c}{HD~58368}\\
\hline
 &    Z  &   solar $log{\epsilon}^a$ & $log{\epsilon}$& [X/H]  & [X/Fe] & $log{\epsilon}$& [X/H]  & [X/Fe]\\
 &       &                         &                  &        &        \\
\hline
Na {\sc i}  &  11     &   6.24   &  6.00$\pm$0.20  &   $-$0.24  &  $-$0.19 &  6.55$\pm$0.15  &    +0.31  &  +0.22  \\
Mg {\sc i}  &  12     &   7.60   &  7.14$\pm$0.07  &   $-$0.46  &  $-$0.41&  7.39$\pm$0.11  &   $-$0.30  &  $-$0.39   \\
Si {\sc i}  &  14     &   7.51   &  6.88$\pm$0.06  &    $-$0.63  &  $-$0.58&  7.17$\pm$0.19  &   $-$0.34  &  $-$0.43\\
Ca {\sc i}  &  20     &   6.34   &  5.69$\pm$0.20  &   $-$0.65  &  $-$0.60&  6.12$\pm$0.12  &   $-$0.22  &  $-$0.31\\
Sc {\sc ii} &  21     &   3.15   &   3.70$\pm$.20  &     +0.45     &  +0.47 &  3.35$\pm$0.20  &   +0.20  &  +0.08\\
Ti {\sc i}  &  22     &   4.95   &  5.07$\pm$0.14  &   +0.12  &  +0.17&  4.82$\pm$0.16  &   $-$0.13  &  $-$0.22\\
Ti {\sc ii} &  22     &   4.95   &        --       &     --    &   --  & 5.08$\pm$0.01  &    +0.13  &  +0.04\\
V {\sc i}   &  23     &   3.93   &  4.10$\pm$0.20   &    0.17 & 0.22&  3.92$\pm$0.20  &   $-$0.01  & $-$0.10\\
Cr {\sc i}  &  24     &   5.64   &  6.03$\pm$0.18   &   +0.39  &  +0.44&  5.41$\pm$0.11  &   $-$0.23  &  $-$0.32\\
Mn {\sc i}  &  25     &   5.43   &  5.23$\pm$0.20   &   $-$0.20  &  $-$0.15&  5.25$\pm$0.20  &   $-$0.18  & $-$0.27\\
Fe {\sc i}  &  26     &   7.50   &  7.45$\pm$0.13  &  $-$0.05& -&  7.59$\pm$0.12  &  +0.09& -\\
Fe {\sc ii}  &  26     &   7.50   &  7.48$\pm$0.30  &  $-$0.02& -&  7.62$\pm$0.110 &  +0.12& -\\
Co {\sc i}  &  27     &   4.99   &  5.21$\pm$0.13  &    +0.22      & +0.27 &  5.04 $\pm$0.17 &    +0.05  & $-$0.06\\
Ni {\sc i}  &  28     &   6.22   &  6.52$\pm$0.18  &   +0.30  &  +0.35&  6.33$\pm$0.13  &   +0.11   &  +0.02\\
Sr {\sc i}  &  38     &   2.87   &  3.79$\pm$0.20  &   +0.92  & +0.97&  4.15$\pm$0.20  &   +1.28   & +1.19\\
Y {\sc ii}  &  39     &   2.21   &  3.50$\pm$0.20  &   +1.29  &  +1.31&  3.45$\pm$0.04  &   +1.24   &  +1.12\\\
Zr {\sc i} &  40     &   2.58   &  3.29 $\pm$0.20   &   +0.71  & +0.73&  3.40$\pm$0.07  &   +0.82  &+0.73 \\
Ba {\sc ii} &  56     &   2.18   &  3.32$\pm$0.20   &   +1.14  & +1.16 &  3.50$\pm$0.20  &   +1.32  & +1.20\\
La {\sc ii} &  57     &   1.10   &  2.60$\pm$0.20  &   +1.50  & +1.52&  2.70$\pm$0.20  &   +1.60  & +1.48\\
Ce {\sc ii} &  58     &   1.58   &  3.24$\pm$0.04  &   +1.66  & +1.68&  2.97$\pm$0.07  &   +1.39  & +1.27\\
Pr {\sc ii} &  59     &   0.72   &  2.42$\pm$0.08  &   +1.70  & +1.72&  1.83$\pm$0.12  &   +1.11  & +0.99\\
Nd {\sc ii} &  60     &   1.42   &  3.00$\pm$0.12  &   +1.58  & +1.60&  2.47$\pm$0.17  &   +1.05  & +0.93\\
Sm {\sc ii} &  62     &   0.96   &        --       &     --    &   --  &  2.09$\pm$0.20  &   +1.13  & +1.01\\
Eu {\sc ii} &  63     &   0.52   &  1.40$\pm$0.20  &   +0.88 & +0.90 &  0.99$\pm$0.20  &   +0.57   & +0.45\\
Dy {\sc ii}  &  66     &   1.10   &      --       &     --    &   --  &  3.17$\pm$0.20 &   +2.07   &  +1.95\\

\hline 
\end{tabular}

$^{a}$ Asplund et al. (2009) \\
\end{table*}
}

{\footnotesize
\begin{table*}
{\bf Table 7: Elemental abundances in HD~119650 and HD~191010}\\
\begin{tabular}{|c|c|c|c|c|c|c|c|c|}
\hline
& \multicolumn{6}{c}{HD~119650} & \multicolumn{1}{c}{HD~191010}\\
\hline
 &    Z  &   solar $log{\epsilon}^a$ & $log{\epsilon}$& [X/H]  & [X/Fe] & $log{\epsilon}$& [X/H]  & [X/Fe]\\
 &       &                         &                  &        &        \\
\hline
Na {\sc i}  &  11     &   6.24   &  6.52$\pm$0.30  &   +0.28  &  +0.24&  6.96$\pm$0.20 &   +0.72  &  0.59  \\
Mg {\sc i}  &  12     &   7.60   &  7.48$\pm$0.07  &   $-$0.12  &  $-$0.16 &  7.79$\pm$0.08 &   +0.19  &  +0.06    \\
Si {\sc i}  &  14     &   7.51   &  7.03$\pm$0.17  &   $-$0.48 &  $-$0.52&  6.99$\pm$0.20 &   $-$0.52 &  $-$0.65\\
Ca {\sc i}  &  20     &   6.34   &  6.08$\pm$0.19  &   $-$0.26  &  $-$0.30 &  6.29$\pm$0.12  &   $-$0.05  &  $-$0.18\\
Sc {\sc ii} &  21     &   3.15   &  3.25$\pm$0.20  &   $-$0.10  &  $-$0.20 &  3.27$\pm$0.20  &   +0.12  & 0.00 \\
Ti {\sc i}  &  22     &   4.95   &  4.97$\pm$0.17  &   $-$0.02  &  $-$0.06&  4.85$\pm$0.15  &   0.10  &  $-$0.03\\
Ti {\sc ii} &  22     &   4.95   &  5.12$\pm$0.18  &   +0.17     & +0.07&  5.30$\pm$0.07  &   +0.35  &  +0.23 \\
V {\sc i}   &  23     &   3.93   &  4.49$\pm$0.20  &  +0.56   & +0.52 &  3.95$\pm$0.20  &   $-$0.02  &  $-$0.15\\
Cr {\sc i}  &  24     &   5.64   &  5.74$\pm$0.21  &   +0.10     &  +0.06&  5.66$\pm$0.19 &   +0.02  &  $-$0.11\\
Mn {\sc i}  &  25     &   5.43   &  5.32$\pm$0.11  &   $-$0.11  &  $-$0.15 &  5.25$\pm$0.20 &   $-$0.18  &  $-$0.21\\
Fe {\sc i}  &  26     &   7.50   &  7.54$\pm$0.20  &  +0.04& - &  7.63$\pm$0.13  &  +0.13& -\\
Fe {\sc ii}  &  26     &   7.50   &  7.60$\pm$0.20  &  +0.10& - &  7.62$\pm$0.12  &  +0.12& -\\
Co {\sc i}  &  27     &   4.99   &  5.18$\pm$0.16  &   +0.19       & +0.15&  4.91$\pm$0.15  &  $-$0.08  & $-$0.21\\
Ni {\sc i}  &  28     &   6.22   &  6.21$\pm$0.13  &   $-$0.01  &  $-$0.05&  6.16$\pm$0.14 &    $-$0.06 &  $-$0.19\\\
Zn {\sc i}  &  30     &   4.56   &  4.99$\pm$0.20  &   +0.43  &  +0.39 &--&--&-- \\
Sr {\sc i}  &  38     &   2.87   &  3.92$\pm$0.20  &   +1.05  & +1.01 &  4.36 $\pm$0.20 &   +1.49   & +1.36\\
Y {\sc ii}  &  39     &   2.21   &  2.99$\pm$0.18  &   +0.78  &  +0.68&  2.54 $\pm$0.15 &   +0.23  &  +0.11\\
Zr {\sc i} &  40     &   2.58   &  3.04$\pm$0.20  &   +0.46  & +0.42 &--&--&--\\
Ba {\sc ii} &  56     &   2.18   &  2.80$\pm$0.20  &  +0.62   & +0.52 &  2.70 $\pm$0.20 &   +0.52    & +0.40\\
La {\sc ii} &  57     &   1.10   &  2.20$\pm$0.20  &   +1.10  & +1.00&  1.64$\pm$0.20  &   +0.54    & +0.42\\
Ce {\sc ii} &  58     &   1.58   &  2.35$\pm$0.10  &   +0.77  & +0.67&  2.16$\pm$0.06  &   +0.58    & +0.46\\
Pr {\sc ii} &  59     &   0.72   &  1.41$\pm$0.20  &   +0.69  & +0.59&  1.40$\pm$0.20  &   +0.68    & +0.52\\
Nd {\sc ii} &  60     &   1.42   &  1.92$\pm$0.09  &   +0.50  & +0.40&  1.66$\pm$0.09  &   +0.24   & +0.12\\
Sm {\sc ii} &  62     &   0.96   &  2.35$\pm$0.13  &   +1.39  & +1.29&  0.90$\pm$0.20  &   $-$0.06    & $-$0.18\\
Eu {\sc ii} &  63     &   0.52   &  1.00$\pm$0.20  &   +0.48  & +0.38&  0.72$\pm$0.20  &   +0.20 & +0.08\\
Dy {\sc ii} &  66     &   1.10   &  2.74$\pm$0.09  &   +1.64   & +1.54&  2.31$\pm$0.20  & +1.21     &   +1.09\\
\hline 
\end{tabular}

$^{a}$ Asplund et al. (2009) \\
\end{table*}
}

{\footnotesize
\begin{table*}
{\bf Table 8: Equivalent widths (in m\r{A}) of lines used for calculation of elemental abundances.}\\ 
\begin{tabular}{ccccccccc}\hline
Wavelength &Id        &E$_{low}$ & log gf &HD 49641      &HD 58368       &HD 119650 	  &HD 191010	  &references\\
(\r{A})    &          & (eV)     &        &              &               &          	  &         	  &        \\
5682.650   &   Na I   &  2.100   & -0.70  & --           &143.4(6.66)    &158.1(6.72)     & ---    	  &1\\      	   
5688.220   &          &  2.100   & -0.40  & 150.6(6.00)  &150.6(6.44)    &154.1(6.32)     &178.1(6.96)    &2 \\  	  
5889.950   &          &  0.000   &  0.10  & 847.9(5.60)  &746.8(6.06)    &913.1(6.12)     &577.3(6.42)    &3	\\ 	   	  
5895.920   &          &  0.000   & -0.20  & 567.7(5.50)  &507.4(6.00)    &610.6(6.03)     &470.8(6.49)    &3\\
4702.990   &  Mg I    &4.350     & -0.67  & 228.1(7.19)  &219.2(7.46)    &227.6(7.53)     &215.6(7.85)    &4\\
5528.400   &          &4.350     & -0.49  & 237.1(7.09)  &218.2(7.30)    &239.9(7.43)     &227.3(7.73)    &4\\	  
5711.088   &          &4.346     & -1.83  &     --       &  --     	 &138.7(7.69)     &    --   	  &4 \\
4947.607   &Si I      &5.082     &  -1.82 &  19.0(6.92)  & 24.4(7.02)   &  22.2(6.87)     & 28.5(6.99)    &2\\
5666.677   &          &5.616  	 &  -1.05 &  22.2(6.83)  &   -- 	&  36.5(7.00) 	  &    -- 	  &2 \\ 
6138.515   &          &5.984  	 &  -1.35 &   --    	 &   -- 	&    --    	  & 44.7(7.68)    &2\\
\hline
\end{tabular}

The numbers in the paranthesis in columns 5-8 give the derived abundances from the respective line.\\
References:   1. Kurucz (1988), 2. Kurucz \& Peytremann (1975), 3. Wiese et al. (1966), 4. Martin et al. (1988a)  \\
{\bf Note:} This table is available in its entirety in online only. A portion is shown here for guidance regarding its form and content.\\
\end{table*}
}
{\footnotesize
\begin{table*}
{\bf Table 9: Observed values for [Fe/H], [ls/Fe], [hs/Fe]  and [hs/ls]}\\
\begin{tabular}{lccccc}
\hline
Star Name & [Fe/H] & [ls/Fe] & [hs/Fe] & [hs/ls]   \\
\hline
HD 49641   &$-$0.05  & 1.33     &1.49     & 0.16   \\
HD 58368   &0.09  & 1.01     &1.18     & 0.17   \\
HD 119650  &0.04  & 0.70     &0.78     & 0.08   \\
HD 191010  &0.13  & 0.74     &0.24  & $-$0.50   \\
\hline
\end{tabular}

\end{table*}
}
\begin{figure}
\centering
\includegraphics[height=8cm,width=8cm]{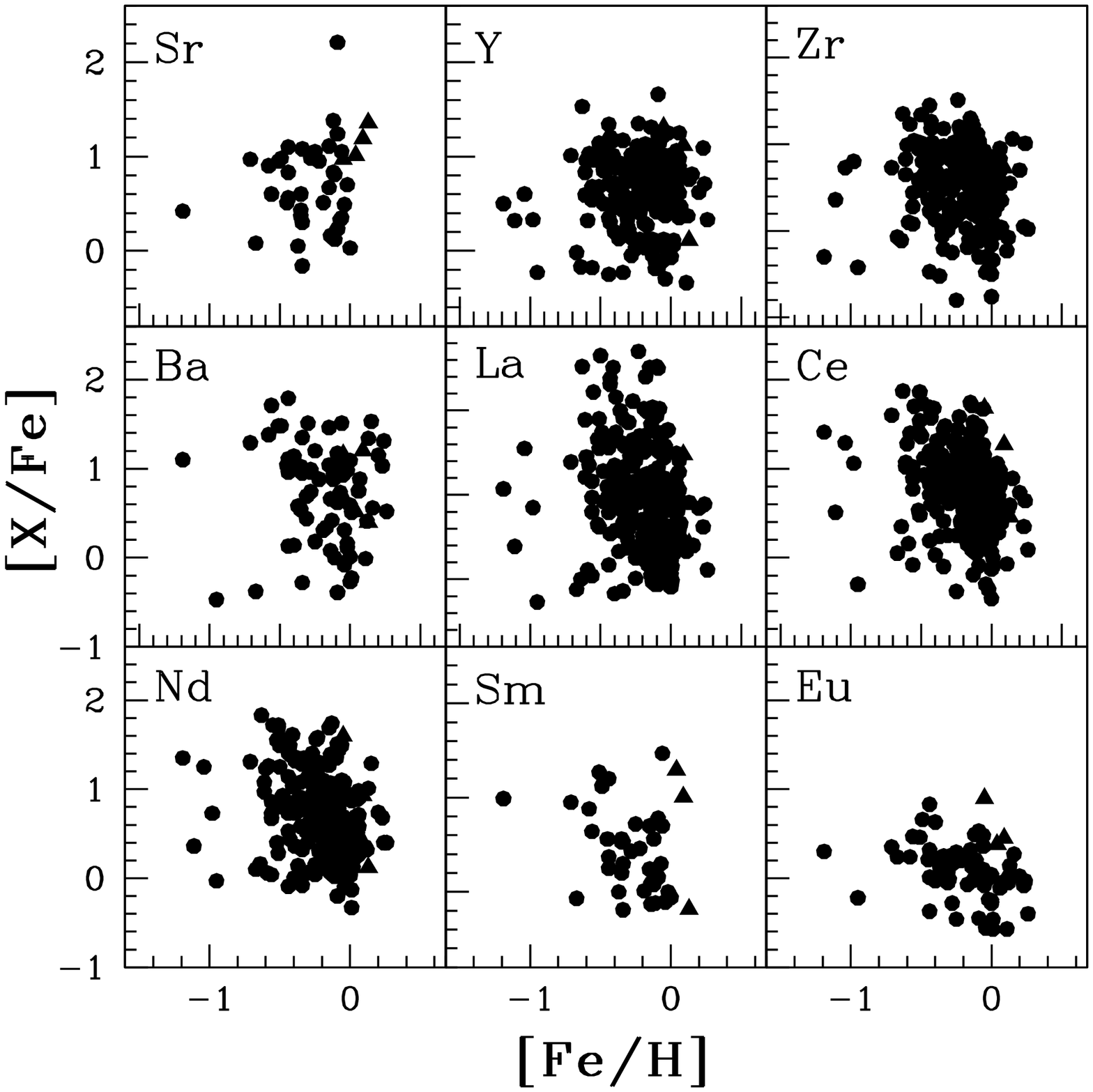}
\includegraphics[height=8cm,width=8cm]{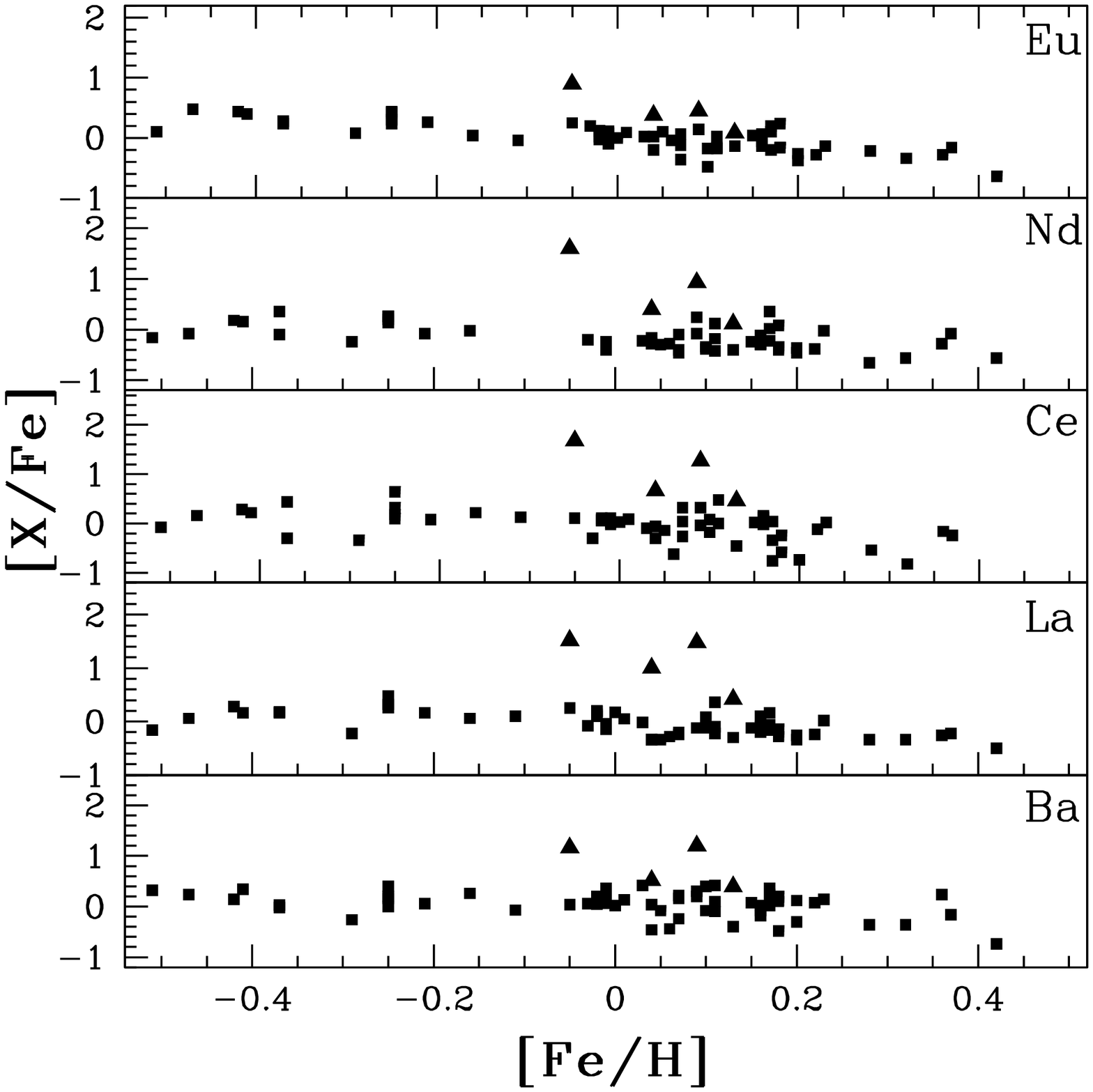}
\caption{Abundance ratios of heavy elements observed in the programme
stars with respect to metallicity [Fe/H]. The abundance ratios show 
a large scatter with respect to metallicity.
In the upper panel our objects are indicated with solid 
triangles. Solid circle represents Ba stars  from literature
 (i.e., Allen \& Barbuy (2006),
 Liang et al. (2003), Smiljanic et al. (2007), Zacs (1994), and 
de Castro et al. (2016).) de Castro et al's data are available only for five
 heavy elements, Y, Zr, La, Ce and Nd.  In the lower panel  solid 
square represents   normal giants from literature (i.e., Van der Swaelmen 
et al. (2016), Tautvaisiene  et al. (2000), Luck and Heiter (2007)); 
our objects are indicated with solid  triangles.}
\end{figure}
\subsection{Stellar Masses}
We have derived the masses of the programme stars from their locations in the 
Hertzsprung-Russel diagram, using the Girardi et al. (2000) database of 
evolutionary tracks and our estimates of log (L/L$_{\odot}$) 
and T$_{eff}$ (figure 7). While  log (L/L$_{\odot}$) is derived by 
photometric 
methods  using the parallax values from the Hipparcos catalogue 
(van Leeuwen 2007); for 
T$_{eff}$ we have used our spectroscopic estimates.  We have 
selected  evolutionary tracks with initial 
compositions, Z = 0.019 and Y = 0.273, where Z is initial metallicity and Y 
is initial helium abundance. The approximated masses derived for the
programme stars  are presented in Table 10. It is to be noted that these 
mass estimates are only indicative as the Hipparcos parallaxes for these 
four objects have large uncertainty in measurements that amounts to 
${\sim}$ 35 to 43 \% for  three stars and  the largest uncertainty 
showing for HD~119650 with ${\sim}$ 75\%.

{\footnotesize
\begin{table*}
{Table 10: \bf {Derived Masses (M$_{\odot}$) of programme stars}}\\
\begin{tabular}{cccc}
\hline
Object    &log (L/L$_{\odot}$) & log(T$_{eff}$)& Mass(in M$_{\odot}$)  \\
          &        &           &            \\ 
\hline
HD~49641   &2.04   & 3.67  &2.5\\
HD~58368   & 1.50  &3.70   &2.3\\
HD~119650  & 1.93  & 3.68  &2.5\\
HD~191010  & 1.96  & 3.73  &3.0\\ 
\hline
\end{tabular}
\end{table*}
}

\begin{figure}
\centering
\includegraphics[height=8cm,width=8cm]{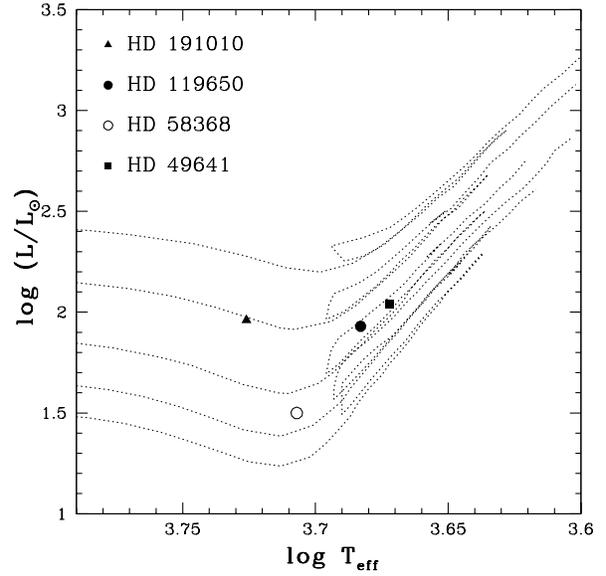}
\caption{ The location of HD~49641, HD~58368, HD~119650 and HD~191010  are 
indicated  in the H-R diagram. The masses are derived using the evolutionary 
tracks of Girardi et al. (2000). The evolutionary tracks are shown for 
masses 2.0, 2.2, 2.5,3.0  and 3.5 M$_{\odot}$ from bottom to top.}
\end{figure}

\begin{table*}
{Table 11: {\bf Best fit coefficients and reduced chi-square values}}\\
\begin{tabular}{lccc}
\hline
Star Name& $A_s $& $A_r$&$ {\chi}^2$\\
\hline
HD~49641 & 0.56 $\pm$ 0.10 & 0.59 $\pm$ 0.09& 1.71\\
HD~58368& 0.67 $\pm$  0.08 & 0.25 $\pm$ 0.08& 0.76\\
HD~119650& 0.39 $\pm$ 0.08 &0.63 $\pm$ 0.07& 0.80\\
HD~191010& 0.80 $\pm$ 0.08 &  0.21 $\pm$ 0.08 & 0.95\\

\hline
\end{tabular}
\end{table*}

\section{Parametric model based study}
We have  
examined  the origin of the neutron-capture elements by comparing the 
observed abundances with the predicted s- and r- process contribution 
using appropriate model function in the framework of a parametric model.

We have utilized the solar system r- and s- process isotopic abundances 
of stellar models offered by Arlandini  et al. (1999). The observed 
elemental abundances are scaled to the metallicity of the corresponding 
 stars and normalized to their respective barium abundances. Elemental 
abundances are then fitted with the parametric model function
$log{\epsilon_i}$  = $A_sN_{is}$  + $ A_r N_{ir}$

where $N_{is}$  indicates the  abundance  
from s-process,   $N_{ir}$ indicates the  abundance  
from r-process; $A_s$ indicates  the component coefficient that 
correspond to   contributions   from  the s-process  and
$A_r$ indicates  the component coefficient that correspond to contributions 
  from the r-process.
  The best fit coefficients 
and reduced chi-square values for the programme stars are given in Table 11.
While in the case of HD~49641 the observed abundances of neutron-capture
elements seem to have similar contribution from s- and r-process, for
the objects HD~58368 and HD~191010 contributions from s-process seem to 
dominate. HD~119650 seem to have more contributions from r-process; this
object also shows Sm and Dy with  values [X/Fe] $>$ 1.0.

{\footnotesize
\begin{table*}
\begin{tiny}
{\bf Table 12: comparison with literature values }\\ 
\begin{tabular}{lcccccccccccccccc}
\hline
starname& T$_{eff}$ & logg & [Fe/H]& [Na/Fe]& [Mg/Fe]& [Si/Fe]& [Ca/Fe]& [Sc/Fe]& [Ti/Fe]& [V/Fe] &[Cr/Fe]& [Mn/Fe]& [Co/Fe]& [Ni/Fe]&   ref\\
\hline
HD49641 &  4700  & 3.4    & -0.05 &  -0.19 &  -0.41 & -0.58   & -0.60 & +0.47  &  +0.17 &  +0.22 & +0.44 &  -0.15 & +0.27  & +0.35  &  1\\
        &  4351  & 1.05   & -0.32 & -0.03 &   0.49 &  0.40  &   0.21 &  0.07  &   0.04 &  -0.02 &  0.17  &  0.20  &  -    &  0.03 &  2\\
        &  4400  & 1.5    & -0.30 & +0.24 &   0.10 &  0.22  &  0.07  &   -   & -0.02  &   -   &  0.04  &  -    &  -   &  0.05 &  3 \\
HD58368 &  5095  & 3.45   &  +0.09 &  +0.22 &  -0.39 & -0.43  &  -0.31 & +0.08  &  -0.22 &  -0.10 & -0.32  & -0.27  &-0.06  & +0.02   &  1\\
        &  4777  & 2.47   & -0.11 &   0.24 &  -0.03 &  0.27  &  -0.08 &  0.09  &  -0.18 &  -0.35 & -0.04  & -0.02  &  -     & 0.02 &  2\\
        &  5000  & 2.6   & +0.04 &  0.11  & -0.13  &  0.12  &  0.06  & -      &  -0.01 &  -     & -0.04  &  -     &  -     & 0.02 & 3  \\
\hline
\end{tabular}
\end{tiny}
1. Our work; 2. Yang et al. (2016); de Castro et al. (2016)\\

\end{table*}
}

 {\footnotesize
\begin{table*}
{\bf Table 12: continued }\\ 
\begin{center}
\begin{tabular}{lcccccccccccc}
\hline
starname& [Sr/Fe]& [Y/Fe]&  [Zr/Fe]& [Ba/Fe]& [La/Fe]& [Ce/Fe]& [Pr/Fe]& [Nd/Fe]& [Sm/Fe]& [Eu/Fe]& [Dy/Fe]& ref  \\
\hline
HD49641 &  0.97 &  +2.29 &   0.73  &  1.16  & 1.52   &  1.68  &   1.72  & 1.60  &   -    &  +0.90 &  -     &  1   \\
         &   -   &   0.35 &  0.45  &  1.13  & 1.38   &    -   &   -    &    -   &   -    &   0.64  & -     &   2\\
         &  -    &   0.89 &  0.53  &   -    & 1.86   &  1.04  &  -     & 1.14   &  -     &   -     & -     &   3 \\
HD58368  & 1.19  &   1.12 &  0.73  &  1.20  & 1.48   &  1.27  &   0.99  &  0.93 &  1.01 &    0.45  &  1.95 &   1 \\
         &  -    &   0.14 &   0.25  &  0.98  & 1.07   &    -   &   -     &   -   &   -    &   0.41 &    -   &   2  \\ 
         &  -    &   0.85  &  0.60  &  -     & 1.13   & 0.86   &  -      &  0.69  &  -     &   -   &    -   &  3  \\
\hline
\end{tabular}
\end{center}
1. Our work; 2. Yang et al. (2016); 3. de Castro et al. (2016)\\
\end{table*}
}

\section{Conclusions}
Elemental abundances are presented  for four peculiar stars listed 
in the Barium star catalogue  of L\"u (1991). All the objects are 
found to be low-velocity objects (V$_{r}<45$ km/s). Except for 
HD~119650 the  estimated radial velocities for these objects are 
signifiacntly different from the previous estimates (Table 3).
 Metallicity estimates indicate that they are mostly solar or 
near-solar  objects, with metallicity [Fe/H] ranging 
from $-$0.03 to $+$0.1 dex. 
Barium is  enhanced in 
HD~49641 and HD~58368 with [Ba/Fe] $>$ 1.0; the other two objects, 
HD~119650 and HD~191010 are found to be mild barium stars with 
  [Ba/Fe] ${\sim}$ 0.52  and 0.4 respectively. 
These two later objects show
large enhancement of Sr, with [Sr/Fe] = 1.01  and 1.36 respectively.

A comparison of  the abundance ratios of the heavy elements with 
those available in literature, including the large sample of de 
Castro et al. (2016) (for Y, Zr, La, Ce and Nd) show  that the
 abundance ratios of the programme stars are well
within   the range generally noticed in  barium stars  (figure 6). 

Anomalies in abundance ratios of ${\alpha}$-elements  are noticed in a 
few objects.  Sc, V and Mn in the 
program stars  show similar  values as  seen in other 
barium stars.  In two objects HD~58368 and HD~119650, Zn shows 
clearly higher values than what 
is generally noticed in barium stars  as well as in normal giants.
 Na, Mg, Si, Sc, V, Cr, and Co
show large scatter with respect to metalllicity. It is to be noted 
that the Galactic Chemical Evolution (GCE) model predictions
for the evolution of Sc, V, and Ti (figure 7, in   Goswami \& 
Prantzos (2000))  show  trends
that are similar to the observed trends of these elements. This implies
that the origin of these elements are likely from massive stars.  
The behavior of Sc is also compatible with 
the results of Nissen et al. (2000), Chen et al. (2000a) and 
Goswami and Prantzos (2000), who demonstrated 
that [Sc/Fe] decreases with increasing metallicity in disk stars. 

The estimated masses  are $2.5$M$_{\odot}$,   $2.3$M$_{\odot}$, 
$2.5$M$_{\odot}$ and $3.0$M$_{\odot}$ for HD~49641, HD~58368, HD~119650 
and HD~191010 respectively.  
The mass of  HD~58368 is within the average mass of 
 typical mild Ba stars, i.e.,  $1.9$  or $2.3$M$_{\odot}$ 
with the $0.60$ or $0.67$ M$_{\odot}$ companion white dwarfs
Jorissen  et al.($1998$). 

It has been argued that a metallicity lower than solar is  
 required to form a barium star. The increasing levels 
of heavy-element overabundances observed in the sequence of mild to 
barium-strong stars support that hypothesis. 
Kovacs ($1985$) also noticed that 
there is a correlation between [Ba/Fe] and metallicity [Fe/H]: strong 
barium stars generally have a metallicity lower than mild barium stars. 
The estimated metallicity of HD~49641 ([Fe/H] = $-$0.03, [Ba/Fe] = 1.16), 
HD~58368 ([Fe/H]= 0.1, [Ba/Fe] = 1.20), 
HD~$119650$ ([Fe/H] = 0.07, [Ba/Fe] = 0.52), 
HD~191010 ([Fe/H] = 0.12, [Ba/Fe] = 0.40) do not exactly show this trend.

The  object HD~49641  clearly shows the characteristics of 
 a barium star with  the heavy elements 
 Ba, La, Ce, Pr and Nd showing  a value [X/Fe] $>$ 1, 
Another interesting feature of this object is its high abundance of Y 
 with [Y/Fe] = 1.31.  With many heavy elements 
significantly enhanced  together with
the variations observed  in radial velocity estimates 
a mass-transfer 
scenario similar to the ones that hold for barium stars, CH stars and 
CEMP-s stars is likely for  the origin of this object. However, 
Europium is also found to be enhanced in this object with [Eu/Fe] = 0.90. 
Such an enhancement of Eu 
can not be explained based on only binary pictures. Further investigation
of its  formation scenario is necessary to understand its abundance patterns. 
 
The object HD~58368  is a  barium star with [Ba/Fe] = 1.20.
Light s-process elements Sr, Y and Zr show large enhancement with
[Sr/Fe] = 1.2, [Y/Fe] = 1.12 and [Zr/Fe] = 0.73. Eu shows a mild
enhancement with [Eu/Fe] = 0.45.  McClure (1983) found this object 
to be a radial velocity variable 
 with a variation of  $\sim \pm $12 Km s$^{-1}$ indicating its binarity. 
A mass-transfer scenario is likely to  hold for this object too. 
 The other  two objects HD~119650, HD~191010 
 show mild enhancement in barium. 
 While La is  enhanced in HD~119650 with [La/Fe] = 1.0, this
element is  mildly enhanced  in HD~191010 with [La/Fe] = 0.42.
Similarly 
Sm is highly abundant in  HD~119650 with [Sm/Fe] = 1.3 and 
underabundant in HD~191010 with [Sm/Fe] = $-$0.18.
 Eu is  only   mildly enhanced in HD~119650 with [Eu/Fe] = 0.38 and 
exhibits a near-solar value  in HD~191010  with  [Eu/Fe] = $+$0.08.
Other elements Ce, Pr and Nd are enhanced but not $>$1.0 with 
respect to Fe.  A mass-transfer scenario suggested for
barium stars might have operated in  these two  objects. 

\vskip 0.3cm
\noindent
{\it Acknowledgements}\\
We are grateful to an anonymous referee for many constructive suggestions
which have improved considerably the readability of the paper.
This work made use of the SIMBAD astronomical data base, operated at CDS, 
Strasbourg, France, and the NASA ADS, USA. We would like to acknowledge  
Sreejith P Babu 
for performing some initial  calculations during his  internship 
 at IIA,  Jan - May, 2014. Funding from the DST project
SB/S2/HEP-010/2013 is greatfully acknowledged. 

{} 

\end{document}